\input{epsf}

\documentclass[journal]{IEEEtran}

\makeatletter
\def\ps@headings{%
\def\@oddhead{\mbox{}\scriptsize\rightmark \hfil \thepage}%
\def\@evenhead{\scriptsize\thepage \hfil \leftmark\mbox{}}%
\def\@oddfoot{}%
\def\@evenfoot{}}

\makeatother
\usepackage[usenames,dvipsnames]{xcolor}
\usepackage{epsf}
\usepackage{graphicx}
\usepackage[cmex10]{amsmath}
\usepackage{amsthm}
\usepackage{amssymb}
\usepackage{epsfig,latexsym,amsmath,epsf,amssymb,amsfonts}
\usepackage{algorithm, algcompatible}
\usepackage{booktabs}       
\usepackage{multirow}

\usepackage{placeins}
\usepackage{url}
\usepackage{cite}
\usepackage{comment}
\usepackage{subfigure}
\usepackage{lipsum, mathtools, cuted}
\usepackage{stfloats}
\usepackage{physics}
\usepackage{color}

\usepackage[top=0.75in, bottom=1in, left=0.625in, right=0.625in]{geometry}
\usepackage{bbm}

\usepackage{amsmath,amsfonts,bm}

\def\Figref#1{Figure~\ref{#1}}

\def\Tableref#1{Table~\ref{#1}}

\def\Secref#1{Section~\ref{#1}}

\def\eqref#1{equation~\ref{#1}}

\def\1{\bm{1}}

\def\rvh{{\mathbf{h}}}

\def\rvv{{\mathbf{v}}}

\def\rvx{{\mathbf{x}}}
\def\rvy{{\mathbf{y}}}

\def\vtheta{{\bm{\theta}}}

\DeclareMathAlphabet{\mathsfit}{\encodingdefault}{\sfdefault}{m}{sl}
\SetMathAlphabet{\mathsfit}{bold}{\encodingdefault}{\sfdefault}{bx}{n}

\def\gC{{\mathcal{C}}}

\def\gL{{\mathcal{L}}}

\def\gX{{\mathcal{X}}}

\def\sR{{\mathbb{R}}}

\newcommand{\MetName}{FedOCS }

\begin{document}
\title{Communication Efficient Distributed Learning over Wireless Channels}
\author{Idan Achituve, Wenbo~Wang~\IEEEmembership{Senior Member,~IEEE}, Ethan Fetaya and
Amir~Leshem,~\IEEEmembership{Fellow,~IEEE}
\thanks{
The authors are with the Faculty of Engineering, Bar-Ilan University,
Ramat Gan 5290002, Israel (e-mails:  idan.achituve@biu.ac.il; \mbox{wangwen}@biu.ac.il; ethanfetaya@gmail.com; amir.leshem@biu.ac.il).
 }\vspace{-5mm}}

\maketitle
\begin{abstract}
Vertical distributed learning exploits the local features collected by multiple learning workers to form a better global model. However, the exchange of data between the workers and the model aggregator for parameter training incurs a heavy communication burden, especially when the learning system is built upon capacity-constrained wireless networks. In this paper, we propose a novel hierarchical distributed learning framework, where each worker separately learns a low-dimensional embedding of their local observed data. Then, they perform communication efficient distributed max-pooling for efficiently transmitting the synthesized input to the aggregator. For data exchange over a shared wireless channel, we propose an opportunistic carrier sensing-based protocol to implement the max-pooling operation for the output data from all the learning workers. Our simulation experiments show that the proposed learning framework is able to achieve almost the same model accuracy as the learning model using the concatenation of all the raw outputs from the learning workers, while requiring a communication load that is independent of the number of workers.
\end{abstract}
\section{Introduction}
\label{sec_introduction}
In a typical vertical distributed or federated learning system (e.g.,~\cite{yang2019federatedlearning,verbraeken2021distributedlearning}) built upon wireless channels, multiple learning parties hold either correlated views or disjoint parts of the same global input data for all samples. For example, in a wireless sensor network for smart surveillance, a number of security cameras with overlapping fields of the same region of interest may work together with several acoustic sensors to identify the behavior pattern of an intruder based on the fusion of both visual and acoustic features. Unlike the traditional centralized approaches of multi-view/multi-modal learning (e.g.,~\cite{poria-etal-2015-deep}), a vertical distributed learning system is expected to perform most of the computations locally at each individual worker, while only sending the minimal necessary information for data/decision fusion to the leader node or the data fusion center. By doing so, the efficiency of data communication between the learning workers and the server can be significantly improved, and better data privacy can be preserved on the worker side.

In the literature, the study of vertical distributed learning can be traced back to the distributed source coding problems, where the resource-constrained wireless sensors collect and encode their local views of the same phenomenon. The sensors then send their local data to a fusion center to exploit the correlated and complementary information of those views for better data reconstruction. Due to the limited communication capability of the wireless channels, data compression is typically needed at the sensor side before relaying the data to the fusion center. This leads to the famous algorithms such as Slepian Wolf~\cite{1055037} and Wyner-Ziv coding~\cite{1055508} and a series of linear projection-based techniques that aim to minimize the distortion of the recovered data due to compression~\cite{4016296, 4957624, 5504834, GISPAN201716}.

With the technological development of Internet of Things (IoT), practical machine learning tasks deployed in IoT networks are frequently conducted based on the vertically partitioned views of the same source entity. Compared to the distributed source coding problems, for vertical distributed learning scenarios a common data label is typically maintained by the fusion center. Moreover, more complex data processing operations (e.g., inference) are performed rather than merely reconstructing the global data at the original dimension. Nowadays, techniques ranging from simple fusion of local learning decisions by the distributed workers~\cite{poria-etal-2015-deep} to model and loss function partition~\cite{9440789, 9042352} are all considered possible options in the investigation of an efficient vertical distributed learning framework.

Designing efficient vertical distributed learning algorithms still needs to address a number of open challenges. In practice, it has been recognized that for general distributed computing tasks, the computation load and the communication load are inversely proportional to each other~\cite{8051074}. As a result, in our context, the goals of short local computation/training time (i.e., efficiency), low inter-node communication load (i.e., scalability), and high learning accuracy cannot be simultaneously achieved. Under the real-world constraint on the inter-node communication capacity, this tradeoff requires the size of the local models at each worker to be properly selected, and the data exchange pattern in the server-worker hierarchy to be carefully designed. For this reason, we propose a novel framework of vertical distributed learning based on max-pooling. More specifically, instead of applying the traditional encoding techniques, such as compression at the raw local data level~\cite{pmlr-v119-li20g} and sparsification at the gradient level~\cite{Li2021infocomSparsification}, we apply the max-pooling operation to combine the intermediate data-representation output by each worker, and employ the opportunistic carrier sensing protocol to reduce the wireless transmission load between the learning workers and the fusion center. Our experiments show that for a learning system of $N$ workers, the proposed learning scheme achieves almost the same inference perforance as the learning algorithm that uses the cancatenated data from all the workers, while consuming only $O(1/N)$ of the total communication load required by collecting the full data from all workers.

\section{Design of the Learning Framework}
\label{sec_system_model}
\subsection{System Model}
\begin{figure}[t!]
  \centering
  \includegraphics[width=0.40\textwidth]{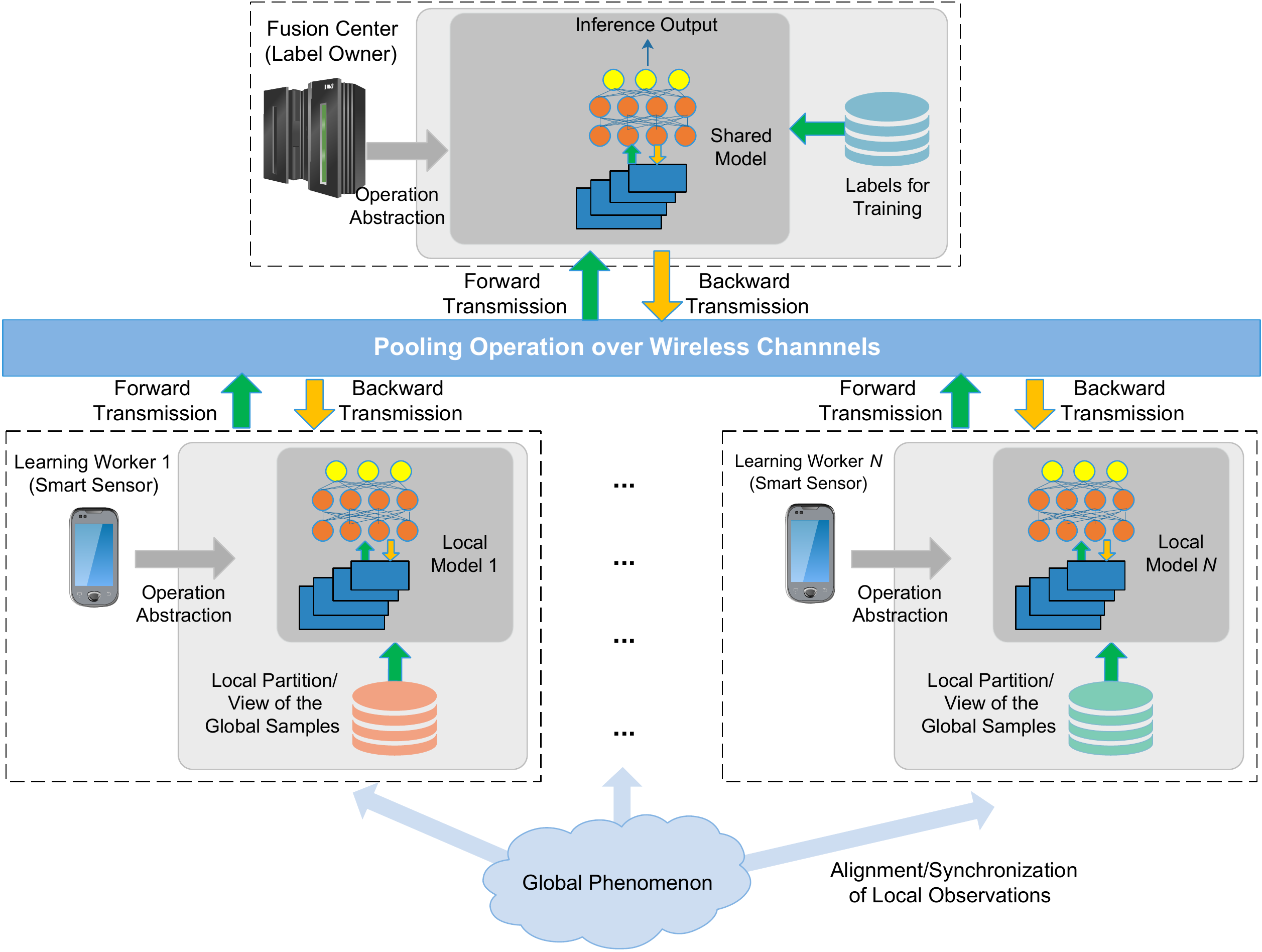}
  \caption{Illustration of the proposed distributed learning framework.}
  \label{fig:model_framework}
\end{figure}

We consider a distributed learning system built upon an IoT network (see Figure~\ref{fig:model_framework}), where $N$ wireless devices (e.g., smart sensors) provide multiple views or sliced-observations/features of the same phenomenon and use a shared wireless channel for inter-node communication. A single fusion center is responsible to synchronize the $N$ distributed learning workers and align their locally observed data to the identical labels. After data alignment, a snapshot of the local raw data collectively provided by the $N$ workers constitutes a global raw feature $\rvx=[\rvx_1^{\textrm{T}},\ldots, \rvx_N^{\textrm{T}}]^{\textrm{T}}$. Instead of directly transmitting the local raw data to the fusion center, worker $n$ trains its local model to learn a low-dimensional embedding $\rvh_n=f_n(\rvx_n; \vtheta_n)$, which is parameterized by $\vtheta_n$. Without loss of generality, assume that worker $n$ deploys a Neural Network (NN) of $L$ layers with a parameter set $\vtheta_n=\{\bm{\omega}_n^\ell\}_{\ell=1}^L$, where $\bm{\omega}_n^\ell$ represents either a convolutional operator or the connection weights (including the biases) to be learned for the $\ell$-th layer. Then, $\rvh_n$ is obtained using the following transformation:
\begin{equation}
  \label{eq_general_transform}
   f_n(\rvx_n; \vtheta_n) \!=\! \left(\bm{\omega}^L_n\right)^{\textrm{T}}\sigma_L\left(\left(\bm{\omega}^{L-1}_n\right)^{\textrm{T}}\sigma_{L-1}\left(...\sigma_1\left( \left(\bm{\omega}^1_n\right)^{\textrm{T}} \rvx_n\right)\right)\right),
\end{equation}
where $\sigma_{\ell}$ is a non-linear operation such as a ReLU activation or a hyperbolic tangent function.

The shared model at the fusion center can be expressed in a similar way to (\ref{eq_general_transform}) with a parameter set $\vtheta_0$ and the concatenated worker outputs $\rvh=[\rvh_{1}^{\textrm{T}},\ldots, \rvh_{N}^{\textrm{T}}]^{\textrm{T}}$ as $\hat{\rvy}=f_0(\rvh; \vtheta_0)$. Let $\rvy$ denote the label that is corresponding to $\rvx$, held by the fusion center. For a given set of training samples and their corresponding labels $\gX=\{(\rvx^{[m]},\rvy^{[m]})\}_{m=1}^{|\gX|}$, the designing goal of the hierarchical framework for distributed learning is to minimize the expectation of the difference between the inference by the fusion center and the labels as follow:
\begin{eqnarray}
  \label{eq_objective}
  (\vtheta^*_0, \ldots, \vtheta^*_N) \!=\!
  \arg\!\min_{\vtheta_0,\ldots, \vtheta_N} \mathop{\mathbb{E}}\nolimits_{\rvx^{[1]},\ldots, \rvx^{|\gX|}}\!\frac{1}{|\gX|}\! \sum_{m=1}^{|\gX|}\gL\! \left(\hat{\rvy}^{[m]}, \rvy^{[m]}  \right),\nonumber
\end{eqnarray}
where $\gL(\cdot)$ is the loss function for the specific learning task at the fusion center\footnote{For simplicity we ignore the index $m$ of the samples in what follows.}.
Then, for regression problems, we can use the following squared error loss for each sample:
\begin{equation}
  \label{squared_loss}
  \gL(\rvy, \hat{\rvy}) = ||\rvy - \hat{\rvy}({\rvh}) ||^2.
\end{equation}
For classification tasks, we can use the cross entropy loss:
\begin{align}
  \label{ce_loss}
  \gL(\rvy, \hat{\rvy}) & = - \sum_{c\in\gC} \mathbf{1}_{\{\rvy = c\}} \log~\Pr(\hat{\rvy}=c |{\rvh}; \vtheta_0), 
\end{align}
where $\mathbf{1}_{\{a = b\}}$ denote the $0/1$ indicator function.

\subsection{Feature Aggregation based on Max Pooling}
\label{learning_with_maxpooling}
Following the system model we now describe the use of max-pooling in order to reducve the features communicated to the fusion center. We assume that each worker observes local $d$-dimensional raw features, i.e., $\rvx_n\in\sR^{d}$ and the output of each worker is a $K$-dimensional feature vector as a low-dimensional description of the local raw feature. Sending all these features over the air would result in transmission of $O(NK)$ messages. To reduce the communication even further, we introduce the max-pooling operation for the fusion center to aggregate the intermediate outputs of all the workers, $\rvh=[\rvh_{1}^{\textrm{T}},\ldots, \rvh_{N}^{\textrm{T}}]^{\textrm{T}}$, as the input of the shared model $f(\cdot;\vtheta)$ at the fusion center. More specifically, max-pooling produces a single $K$-dimensional feature, denoted by $\rvv=[v_1,\ldots, v_K]^{\textrm{T}}$, out of the $N$ local features fed in by the workers, with
\begin{equation}
  \label{eq_max_pooling}
  v_k=\max_{n\in\{1,\ldots, N\}} \{\rvh_{1, k}, \ldots, \rvh_{N, k} \}, k=1,...,K.
\end{equation}
Observing (\ref{eq_max_pooling}), we note that it is possible for a worker $n$ to send only its winning elements in $\rvh_{n}$ for max-pooling to the fusion center, instead of sending the entire output feature vector. Namely, for each $k$ only a single worker transmits a feature element $k$. By doing so, the forward communication load from the workers to the server can be reduced from $O(NK)$ to $O(K)$ in terms of the total dimension of the output features on the worker side. In~\Secref{sec_protocol_design}, we will present a distributed mechanism which implements the pooling operation by providing access to the channel only to the worker holding the highest value of the same feature element.

The max-pooling operation can also save the communication cost from the fusion center to the workers during the process of gradient backpropagation (i.e., backward transmission in Figure~\ref{fig:model_framework}). Following the standard backpropagation procedure, for the $n$-th worker ($n=1,\ldots, N$) we have
\begin{equation}
  \label{eq_back_prop_max_pooling}
  \frac{\partial \gL}{\partial\vtheta_n}=\frac{\partial \gL}{\partial\rvv}\frac{\partial \rvv}{\partial\rvh_n}\frac{\partial \rvh_n}{\partial\vtheta_n}.
\end{equation}
By (\ref{eq_max_pooling}), $\forall k=1,\ldots, K$ we have $\frac{\partial \rvv}{\partial\rvh_n}\in\mathbb{R}^{K\times K}$ and
\begin{equation}
  \label{back_prop_max_pooling}
  \left[\frac{\partial \rvv}{\partial\rvh_n}\right]_{j,k}=\left\{
  \begin{array}{ll}
    1, & \textrm{if } n=\arg\max \{\rvh_{1, k}, \ldots, \rvh_{N, k} \}, j=k\\
    0, & {\textrm{otherwise}}.
  \end{array}\right.
\end{equation}
Therefore, the fusion center needs only to broadcast the vector $\frac{\partial \gL}{\partial\rvv}$ once to all the workers or send only the elements to the corresponding winning workers according to (\ref{eq_max_pooling}).

\section{Data Exchange Protocol using Opportunistic Carrier Sensing}
\label{sec_protocol_design}
To reduce the communication loads needed for computing (\ref{eq_max_pooling}), we apply Opportunistic Carrier Sensing (OCS~\cite{zhao2005opportunistic}) over the shared wireless channel for max-pooling of the $N$ embedding features. Note that when multiple channels are available as in OFDMA, the multi-channel opportunistic carrier-sensing can be applied in a similar manner to~\cite{6117764}. The key idea of OCS is to map the $k$-th element of a feature into each worker's backoff strategy in the corresponding sub-frame of contention resolution. Then, only the first node completing the backoff needs to send the corresponding feature element to the server in the subsequent data transmission sub-frame. Based on (\ref{eq_max_pooling}), we consider that worker $n$ chooses a backoff period $t_{n,k}=g(\rvh_{n,k})$ for the $k$-th element of its output feature. Here, $g(z)$ is a predetermined, strictly decreasing function in $z$. Without loss of generality, we consider a non-negative feature element encoded in a binary floating-point format in a length of $D$ bits following the IEEE 745 standard with a fixed number of fraction bits and precision~\cite{7271015}. Then, the inequality of two encoded values can be determined through bit-wise comparison\footnote{When they have different signs, the comparison is done over the single sign bit. Otherwise, the problem reduces to comparing the negative values of them.}. We define the backoff function for $\rvh_{n,k}$ as follows:
\begin{equation}
  \label{eq_back_off}
  g(\rvh_{n,k})=2^D-{\textrm{INT}}(\rvh_{n,k}),
\end{equation}
where the operation ${\textrm{INT}}(\cdot)$ forces a floating-point value $\rvh_{n,k}$ into the integer expression.

Using (\ref{eq_back_off}), any output feature element of worker $n$ can be quantized into $D$ time slots. As indicated in Algorithm~\ref{alg_ocs}, the max-pooling protocol is composed of $K$ sub-frames, each of which corresponds to one element in the features to be aggregated. A sub-frame contains $D$ sub-slots for backoff contention. More specifically, as described in Line 2 of Algorithm~\ref{alg_ocs}, the operation $\textrm{BIN}_d(g(\rvh_{n,k}))$ ($1\le d\le D$) outputs the binary value of $g(\rvh_{n,k})$ at the $d$-th digit and determines whether an worker $n$ needs to backoff (if $\textrm{BIN}_d(g(\rvh_{n,k}))=1$) at sub-slot $d$ of sub-frame $k$. The workers who sense other transmissions (blocking signals) before their own transmission will stop their contention in the future sub-slots. The workers who are able to transmit the blocking signal but do not receive the ACK signal from the fusion center continue their contention in the next sub-slot.

\begin{algorithm}[t]
 \begin{small}
   \caption{Max-pooling operation through OCS by worker $n$ (in parallel) at sub-frame $k=1,\ldots, K$.}
   \begin{algorithmic}[1]
        \FORALL {$d=1,\ldots, D$}
            \STATE Sense after the backoff in ${\textrm{BIN}}_d(g(\rvh_{n,k}))\in\{0, 1\}$ sub-slot

            \IF {a blocking signal is detected}
                 \STATE {\bf{exit}} \hfill\COMMENT{quit the contention as a loser}
            \ELSE
                \IF {$\textrm{ACK}=0$} \hfill\COMMENT{No ACK is received}
                    \STATE {\bf{continue}} \hfill\COMMENT{continue the contention in sub-slot $d+1$}
                \ELSIF {$\textrm{ACK}=n$}
                    \STATE Transmits $\rvh_{n,k}$ to the server \hfill\COMMENT{winning the contention}
                \ELSE \hfill\COMMENT{some other worker receives the ACK}
                    \STATE {\bf{exit}} \hfill\COMMENT{quit the contention as a loser}
                \ENDIF
            \ENDIF
        \ENDFOR
   \end{algorithmic}
   \label{alg_ocs}
 \end{small}
\end{algorithm}

\section{Simulation Results}
\label{sec:simulation}
We evaluate the model performance of our proposed distributed learning framework in two different learning scenarios. The first is a noisy reconstruction task on the MNIST dataset~\cite{lecun1998gradient}, and the second is a discrimnative task on the CIFAR-10 and CIFAR-100 datasets~\cite{krizhevsky2009learning} which are widely used in federated learning (e.g., \cite{achituve2021personalized} that is built on \cite{achituve2021gp}). MNIST is a dataset of $28 \times 28$ grayscale images of handwritten digits, while the CIFAR-10 and CIFAR-100 datasets contain $32 \times 32$ RGB images from 10 and 100 distinct classes, respectively. The $70$K images contained in MNIST are partitioned into a training set of $60$K images and a testing set with the rest samples. Similarly, CIFAR-10 and CIFAR-100 are partitioned into a training set of $50$K samples and a testing set of $10$K samples. We allocate $10\%$ images from the training set of each dataset for a validation set, which is used for hyper-parameter tunning and validation-based early stopping. Since the opportunistic carrier sensing method results in a single transmission per sub-frame, we assume that optimal coding is applied and error free transmission is achieved.

\begin{figure}[t]
    \centering
    \includegraphics[width=0.8\linewidth]{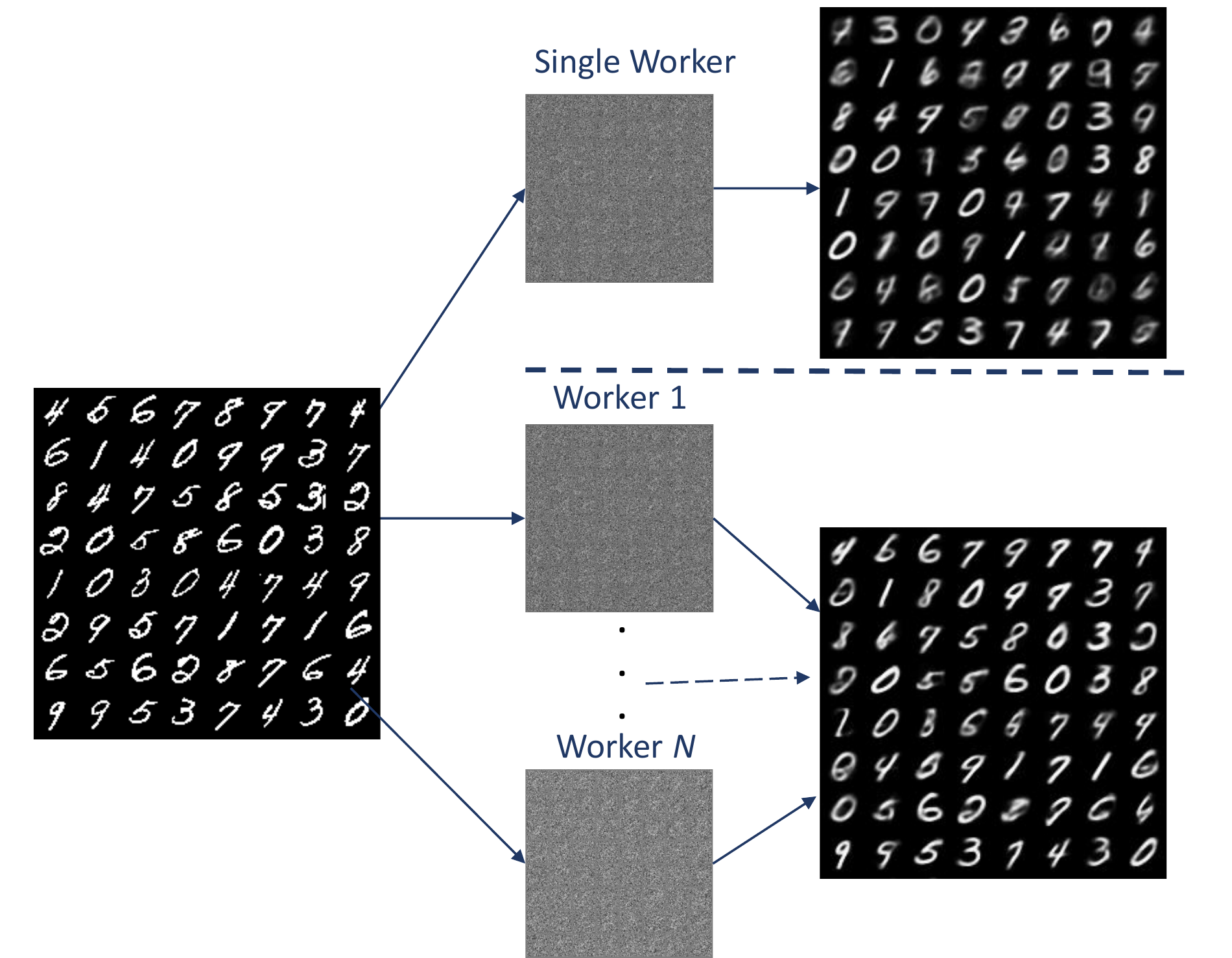}
    \caption{
    Reconstruction of $64$ \textit{test} images, each of which is perturbed by an additive Gaussian noise. The left column shows the original input images arranged in an $8 \cross 8$ grid. The middle column shows the distorted images (views) that each worker samples. The right column shows the reconstruction based on the learning results of one worker (top) and $N=4$ workers (bottom).
    }
    \label{fig:recon}
\end{figure}

\subsection{Handwritten Digit Reconstruction}
One common scenario of vertical distributed learning with shared labels can be found in the reconstruction of the source data from multiple sensors, which may experience different lossy sampling processes. This implies that each senor has a different distortion of the (same) signal. To simulate how our method works in this scenario, we generate different local views of the wireless sensors based on the synthetic data using the MNIST dataset. We consider that a source image is sampled independently by $N=4$ sensors. Each local sample of a sensor is scaled to the range $[0,1]$ and corrupted by an independent observation noise drawn from a Gaussian distribution with a fixed standard deviation of $\sigma = 2$ (see Fig. \ref{fig:recon} for example of noisy images). The model aggregator reconstructs the clean image from the local image representation learned by the sensors. Clearly, training the model requires to leverage the information from all workers. We refer to the NN on each worker as an ``encoder'' and the NN on the model aggregator as the ``decoder''. Each encoder samples a flattened view of a noisy MNIST image of size $784$  and then transforms it to an embedding vector of size $64$ using the local NN. The decoder reconstructs the image of the original dimension from the low-dimensional embedding vectors output by the workers. We deploy small-size networks for both the encoders and decoder and adopt a bottleneck architecture between them. The encoders adopt an NN of three hidden layers with the same sizes $\{512, 256, 128\}$. Similarly, the decoder adopts an NN of three hidden layers with sizes $\{128, 256, 512\}$. \Figref{fig:recon} depicts the reconstruction results of $64$ images in a system with one worker as opposed to that with 4 workers. The quality of reconstruction in the latter system is significantly better even though the two systems communicate the same number of parameters over the wireless channel. Quantitatively, the average negative log-likelihood of the system with 4 workers is $0.13$ as opposed to that with one worker which is $0.19$.

\subsection{Prediction using Patches of a Global Image}
\setlength{\tabcolsep}{2pt}
\begin{table}[!t]
\centering
\caption{ Test accuracy ($\pm$ std) with 4 and 9 works on CIFAR-10 and CIFAR-100, respectively. The numbers in bold are the statistically significant best results.}
\scalebox{1.}{
    \scriptsize
    \begin{tabular}{l c cc cc c cc cc}
    \toprule
    \multirow{2}{*}{}
    && \multicolumn{3}{c}{CIFAR-10} && \multicolumn{3}{c}{CIFAR-100}\\
    \cmidrule(l){2-5}  \cmidrule(l){6-9}
    ~\quad\quad No. of clients &&4 &&9 &&4 &&9\\
    \midrule
    Concat Workers Embed && 84.35 $\pm$ 0.5 && 81.02 $\pm$ 0.1 && 53.06 $\pm$ 0.2 && 51.05 $\pm$ 0.5 \\
    \hline
    Best Worker Pred && 59.31 $\pm$ 0.6 && 44.87 $\pm$ 1.3 && 18.67 $\pm$ 1.4 && 7.97 $\pm$ 1.6\\
    Avg. Workers Preds && 76.03 $\pm$ 0.6 && 61.60 $\pm$ 0.6 && 31.14 $\pm$ 0.7 && 16.89 $\pm$ 1.5\\
    Avg. Workers Embed && 84.28 $\pm$ 0.2 && 80.58 $\pm$ 0.5 && 50.80 $\pm$ 0.3 && 45.78 $\pm$ 0.3\\
    \MetName && 84.48 $\pm$ 0.7 && 80.52 $\pm$ 0.3 && \textbf{53.98 $\pm$ 0.7} && \textbf{49.79 $\pm$ 0.2}\\
    \bottomrule
    \end{tabular}
}
\label{tab:1}
\end{table}

In our second experiment, we consider a scenario where the wireless sensors sample disjoint parts of the same global data. To simulate this scenario, we partition the images from CIFAR-10 and CIFAR-100 to a grid of $4$ and $9$ cells and assign to each worker ($4$ and $9$ for the two datasets, respectively) a fixed unique cell to observe. Each sensing worker uses an NN to generate a low-dimensional embedding representation of its locally observed cell. The fusion center then aggregates the embeddings of all clients using a different NN ("classification head") to learn a distribution over classes. We used MobileNetV2~\cite{sandler2018mobilenetv2} as a feature extractor on each worker, since it balances well between good model performance and low computational requirement, e.g., in terms of memory consumption and training time. At the fusion center, we adopt a fully connected NN with 3 hidden layers of sizes $\{512, 512, 512\}$.

To evaluate the efficiency of our proposed learning framework, we compared the performance of the following methods:
\begin{itemize}
  \item [(1)] \textbf{Avg. Workers Preds}: Each worker generates a separate class distribution. The fusion center then averages over all the local probability vectors to obtain a global distribution over the classes. Note that in this case, each worker has a separate classification head as well.
  \item [(2)] \textbf{Best Worker Pred}: The model of the best performance among the independent local models in (1).
  \item [(3)] \textbf{Avg. Workers Embed}: An alternative pooling method that averages over each dimension of the embedding output across all the workers instead of taking the maximum. The new representation is used as the input to the classification head at the server. {This method requires each worker to send its full embedding feature to the server.}
  \item [(4)] \textbf{Concat Workers Embed}:  The outputs of all the workers are concatenated to form an input to the classification head. This baseline is reminiscent of the vertical federated learning methods~\cite{yang2019federatedlearning} with excessive communication loads. It also requires a larger classification head on the model aggregator to accommodate the larger input dimension.
  \item [(5)] \textbf{FedOCS}: Our proposed method uses max-pooling for feature aggreation as the input to the classification head.
\end{itemize}
The results are presented in \Tableref{tab:1}, where we report the accuracy and the standard deviation based on three runs with random seeds. From the table, we notice that aggregating information in the feature space is considerably more effective compared to the individual baselines in which each client makes a separate decision. In addition, aside from the communication advantage of our method compared to the baseline that performs mean-pooling, \MetName manages to maintain the same accuracy on CIFAR-10 and even surpass it on the harder classification task of CIFAR-100 images. Finally, \MetName usually performs similar to or slightly below the baseline which concatenates the embedding representation. This baseline can be considered as an upper-bound for \MetName performance (yet requires a much larger communication overhead).

\section{Conclusion}
\label{sec_conclusion}
In this paper, we considered a scenario of distributed learning in wireless environments. We proposed a novel hierarchical learning framework for vertical distributed learning tasks with different local views/parts sharing a common label of the global data. Each worker applies an independent neural network model generate a low-dimensional representation of its locally sampled data. We introduced an elegant solution to achieve high communication efficiency through the max-pooling operation which aggregates the intermediate embedding output of these workers based on opportunistic carrier sensing. The fusion center uses the aggregated feature to produce the final output of the learning tasks. 
Our proposed method significantly reduces the bidirectional transmission load between the workers and the fusion center. Simulation results demonstrated that our proposed method can achieve almost the same accuracy as transmitting all the features from all workers and concatenating it at the fusion center, with a communication load that is linear in the size of the local output features and independent of the number of workers.


\bibliographystyle{IEEEtran}
\bibliography{ref}

\end{document}